\documentclass[english,aps,jap,twocolumn,floatfix]{revtex4-1}

\usepackage[T1]{fontenc}
\usepackage[latin9]{inputenc}
\usepackage[active]{srcltx}
\usepackage{booktabs}
\usepackage{units}
\usepackage{textcomp}
\usepackage{amstext}
\usepackage{amssymb}
\usepackage{graphicx}

\makeatletter

\providecommand{\tabularnewline}{\\}

\usepackage{epstopdf}

\makeatother

\begin{document}

\title{Suppression of Spin Pumping Between Ni$_{80}$Fe$_{20}$ and Cu by
a Graphene Interlayer}

\author{Will Gannett}

\author{Mark W. Keller}

\email{mark.keller@nist.gov}

\author{Hans T. Nembach}

\author{Thomas J. Silva}

\author{Ann Chiaramonti Debay}

\affiliation{National Institute of Standards and Technology, Boulder, CO 80305\bigskip{}
}
\begin{abstract}
We compare ferromagnetic resonance measurements of Permalloy Ni$_{80}$Fe$_{20}$
(Py) films sputtered onto Cu(111) films with and without a graphene
(Gr) interlayer grown by chemical vapor deposition before Py deposition.
A two-angle sputtering method ensured that neither Gr nor Py was degraded
by the sample preparation process. We find the expected damping enhancement
from spin pumping for the Py/Cu case and no detectable enhancement
for the Py/Gr/Cu case. Since damping is sensitive to effects other
than spin pumping, we used magnetometry to verify that differences
in Py magnetostatic properties are not responsible for the difference
in damping. We attribute the suppression of spin pumping in Py/Gr/Cu
to the large contact resistance of the Gr/Cu interface.
\end{abstract}

\thanks{Contribution of the National Institute of Standards and Technology;
not subject to copyright in the United States.}

\maketitle

\section{Introduction}

The complex interactions between magnetization dynamics in a ferromagnet
and the flow of charge and spin currents is a topic that combines
rich physics and relevance to technological applications. These interactions
are often divided into two types that are reciprocal manifestations
of the same microscopic process \citep{Brataas:2014ys}. \emph{Spin
transfer} refers to the torque exerted on the magnetization by a flow
of spins (with or without an associated flow of charge). \emph{Spin
pumping} refers to the flow of spins generated by a precessing magnetization.
The first effect enables control of magnetization through an applied
current \citep{Ralph:2008pr}, rather than an applied magnetic field,
and is crucial to realizing a scalable magnetic random access memory
\citep{Ando:2014kl}. The second effect is most often encountered
as an enhanced damping of magnetization dynamics when a thin ferromagnetic
(FM) film is in contact with a nonmagnetic (NM) material that absorbs
the spin current \citep{Tserkovnyak:2002wp,Tserkovnyak:2002tg}. This
loss of angular momentum adds to the intrinsic dissipation of the
FM material, causing an enhancement of the Gilbert damping parameter
$\alpha$. The effect on damping can be quite large $\left(\gtrsim2\times\right)$
for ultrathin $\left(\lesssim\unit[10]{nm}\right)$ FM films and has
been observed for many FM/NM combinations \citep{Tserkovnyak:2002tg,Xia:2002dk,Bauer:2003dz,Gerrits:2006qy,Shaw:2012wu,Mizukami:2001oq,Mizukami:2001ud}.

Brataas \emph{et al.} \citep{Brataas:2014ys} have reviewed the phenomenology
and the quantitative theory of spin transfer and spin pumping. When
considering the effect of spin pumping on FM damping, it is useful
to identify two limiting cases. If the FM is in contact with a material
that strongly scatters or absorbs the pumped spin current, $\alpha$
is increased substantially. In contrast, if the NM does not act as
a spin sink, then pumped spins will accumulate in the NM and drive
a diffusive spin current back toward the FM that cancels the spin
pumping current in steady state. In this ``spin battery'' limit,
there is no effect on $\alpha$. For a general case, the change in
damping for a FM film of thickness $d_{\textrm{FM}}$ is determined
by an effective spin conductance per unit area, $G_{\textrm{eff}}$
(units of $\unit{\textrm{\ensuremath{\Omega}}^{-1}\cdot m^{-2}}$),
according to

\begin{equation}
\Delta\alpha_{\textrm{sp}}=\frac{|\gamma|\hbar^{2}}{2e^{2}M_{\textrm{s}}d_{\textrm{FM}}}G_{\textrm{eff}},\label{eq:damping_Geff}
\end{equation}
where $\gamma$ is the gyromagnetic ratio, $e$ is the electron charge,
$\hbar$ is the reduced Planck constant, and $M_{\textrm{s}}$ is
the saturation magnetization of the FM. The conductance $G_{\textrm{eff}}$
includes contributions from the FM/NM interface and from spin diffusion
in the NM material (see Appendix~A).

Graphene (Gr) is expected to be a useful material for spintronics
because its high carrier mobility and low spin scattering should allow
propagation of spin currents over long distances $\left(\gg\unit[1]{\mu m}\right)$
at room temperature. Spin coherence lengths of about $\unit[1]{\mu m}$
have been demonstrated by injecting spin-polarized charge current
from FM contacts into a Gr channel \citep{Han:2011kx}. The main challenge
facing these devices is the limited efficiency for injecting spin-polarized
carriers from the metallic FM into a channel having much lower electrical
conductivity. Even with tunnel barriers between the FM and the Gr
designed to minimize this effect, the efficiency is only about 30\%
\citep{Han:2011kx}. Since spin pumping involves only spins and not
charges flowing between the FM and NM materials, it provides a way
to inject spin current from FM metals into Gr that is not limited
by the mismatch in electrical conductivity.

This paper compares ferromagnetic resonance (FMR) measurements of
Permalloy Ni$_{80}$Fe$_{20}$ (Py) films with nominal thicknesses
between 8~nm and 45~nm sputtered onto Cu(111) films with and without
a Gr interlayer grown by chemical vapor deposition (CVD) before Py
deposition. The Py films were deposited using a two-step process (see
section \ref{sub:Two-angle}) that avoids damage to the Gr and yields
high quality Py. The FMR results show the expected damping enhancement
for the Py/Cu case but no enhancement for the Py/Gr/Cu case. This
unexpectedly strong effect on spin tranport by a monolayer of material
with weak intrinsic spin scattering requires careful validation. Since
FM damping is sensitive to effects other than spin pumping, much of
the paper is concerned with ensuring that neither Gr nor Py is significantly
degraded by the sample preparation process. In particular, we use
a novel two-angle sputtering geometry, we verify the Gr quality using
Raman spectroscopy, we use magnetometry to measure the moment and
coercivity of each sample, and we examine the Py morphology using
transmission electron microscopy.

Previous studies have shown evidence for spin pumping from Py into
Gr \citep{Tang:2013zr,Singh:2015fe}, but the use of only a single
thickness of Py and the lack of magnetometry or other characterization
of the Py itself limits the conclusions that can be drawn from the
observed effects. We discuss these limitations in detail, drawing
on our results and on other recently published work \citep{Berger:2014xe}.
Finally, we attribute the suppression of spin pumping in Py/Gr/Cu
to the large contact resistance of the Gr/Cu interface \citep{Robinson:2011uq}.

\section{FMR Measurement of Spin Pumping}

For FMR measurements, samples were placed on a coplanar waveguide
with a center conductor width of $\unit[100]{\mu m}$, and the scattering
parameter $S_{21}$ was measured using a vector network analyzer.
The static magnetic field $H$ was applied perpendicular to the sample
plane and was large enough to saturate the magnetization, suppressing
the 2-magnon scattering contribution to the FMR linewidth \citep{McMichael:2004mz}.
For each excitation frequency $f$, fitting the real and imaginary
parts of $S_{21}(H)$ yields the resonance field $H_{\textrm{res}}(f)$
and the field-swept linewidth $\Delta H(f)$. Details of the measurement
and data analysis techniques are described in \citep{Nembach:2011uq}.

The FMR resonance field and linewidth as a function of excitation
frequency are given by 

\begin{equation}
H_{\textrm{res}}(f)=\frac{2\pi f}{|\gamma|\mu_{0}}+M_{\textrm{eff}}\label{eq:H_resonance}
\end{equation}

\begin{equation}
\Delta H(f)=\frac{4\pi f\alpha}{|\gamma|\mu_{0}}+\Delta H_{0},\label{eq:Delta_H}
\end{equation}
where $\mu_{0}$ is the magnetic constant, $M_{\textrm{eff}}=M_{\textrm{s}}-H_{\textrm{k}}^{\bot}$
is the effective magnetization including a perpendicular anisotropy
field $H_{\textrm{k}}^{\bot}$, and $\Delta H_{0}$ is the broadening
due to inhomogeneity in the local resonance field across the sample
\citep{Nembach:2011uq}. A linear fit to Eq.~\ref{eq:H_resonance}
gives values for $\gamma$ and $M_{\textrm{eff}}$, and a linear fit
to Eq.~\ref{eq:Delta_H} gives values for $\alpha$ and $\Delta H_{0}$
\citep{Nembach:2011uq}. Results for all four fitted parameters are
given in Appendix~B.

Figure \ref{DampingVsThickness} shows $\alpha$ as a function of
$d_{\textrm{FM}}$ for two sets of Py samples. The Py/Cu samples clearly
show an increase in $\alpha$ as $d_{\textrm{FM}}$ decreases, as
seen previously \citep{Gerrits:2006qy} and as expected from Eq.~\ref{eq:damping_Geff},
while the Py/Gr/Cu samples show no significant change in $\alpha$.
As shown in the next section, the Py in the two sets of samples has
very similar magnetic and morphological properties, so this result
is strong evidence for suppression of spin pumping between Py and
Cu due to the Gr interlayer.

\begin{figure}[!th]
\centering{}\includegraphics[width=8.5cm]{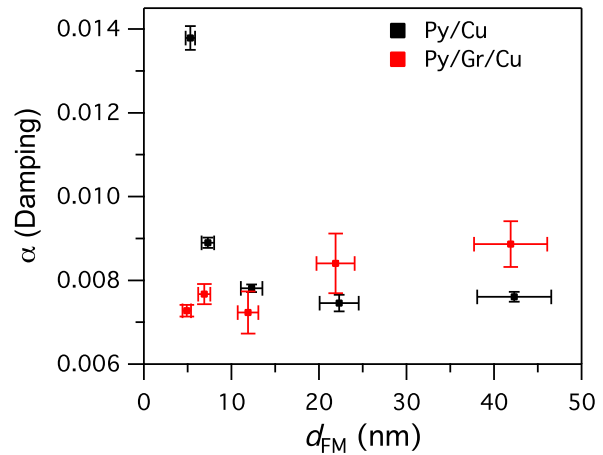}\protect\caption{Damping vs. thickness for Py deposited on bare Cu (Py/Cu) and on Cu
covered by CVD Gr (Py/Gr/Cu). Vertical error bars show the uncertainty
from the linear fits to Eq.~\ref{eq:Delta_H}. The larger uncertainty
in $\alpha$ for the thicker Gr samples is due to a larger scatter
in $\Delta H(f)$ for these samples. Horizontal error bars reflect
an estimated 10\% uncertainty in deposition rate. \label{DampingVsThickness}}
\end{figure}

\section{Sample Preparation and Characterization}

\subsection{Motivation for Two-angle Sputtering \label{sub:Two-angle}}

\begin{figure}[!h]
\begin{centering}
\includegraphics[width=8.5cm]{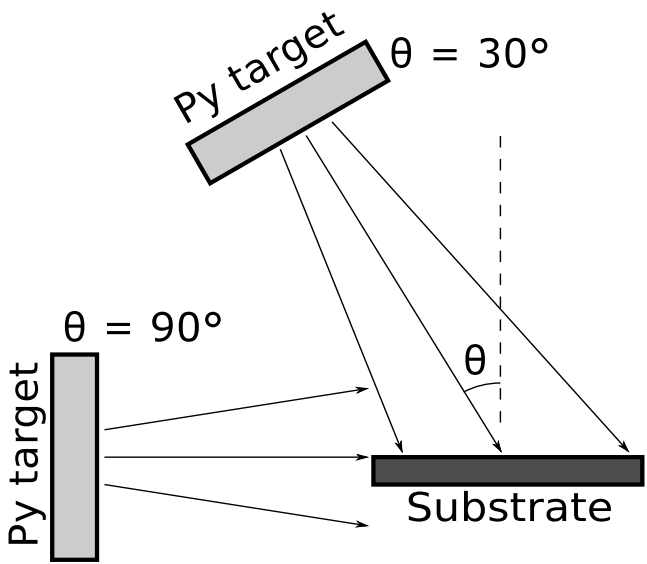}
\par\end{centering}

\centering{}\protect\caption{Schematic of deposition at two angles $\theta$ relative to substrate
normal direction (not to scale). The Py target is shown in different
positions for clarity, but in practice $\theta$ was changed by tilting
the substrate. \label{DepSchematic}}
\end{figure}

Damage to Gr during magnetron sputtering of Ti and Al has been studied
using Gr flakes exfoliated onto Si substrates \citep{Chen:2013ve}.
This work considered several types of bombardment present in sputtering
(electrons, photons, depositing atoms, and inert gas atoms) and concluded
that damage is primarily due to inert gas atoms, which can be reflected
from the target as neutral atoms with sufficient energy to displace
C atoms from Gr. Detailed simulations \citep{Yamamura:1995nx} of
sputtering a Cu target in Ar gas show, for a typical discharge voltage
of 400~V, that Cu atoms leave the target with an average energy of
about 10~eV, while Ar neutrals are reflected with an average energy
of about 45~eV. For comparison, the displacement threshold for Ar
striking Gr at normal incidence is 33~eV \citep{Lehtinen:2010rm,Lehtinen:2011kl}.
Increasing the gas pressure or target-to-substrate distance reduces
the flux of both Ar neutrals and depositing atoms, but since the more
energetic Ar atoms have a longer mean free path than the depositing
atoms \citep{Yamamura:1995nx}, the deposition rate is strongly suppressed
before the Gr is protected from Ar impacts. Since the displacement
threshold increases away from normal incidence \citep{Lehtinen:2010rm,Lehtinen:2011kl},
damage can be avoided by orienting the substrate perpendicular to
the target, as seen in Fig. \ref{DepSchematic} with $\theta=90^{\circ}$.
The tradeoff between deposition rate and Gr damage for orientations
near perpendicular was quantified for Ti and Al in \citep{Chen:2013ve}.

For many magnetic materials, including Py, bombardment of the substrate
is important in promoting growth of thin films that are homogeneous
and have properties such as $M_{\textrm{s}}$ and $\alpha$ that are
close to bulk values. Thus a substrate orientation perpendicular to
the target is undesirable in terms of film quality as well as deposition
rate. For this reason, we chose to deposit Py at two separate orientations,
as shown in Fig.~\ref{DepSchematic}. The first deposition, 5~nm
at $\theta=90^{\circ}$, covers the Gr with a Py film of low quality
but does not damage the Gr, and this Py layer protects the Gr from
damage during the second deposition, at $\theta=30^{\circ}$, that
yields high quality Py. The two depositions were done without breaking
vacuum by use of an angled sample holder (see Methods section). Importantly,
transmission electron microscopy (see Section \ref{sub:Py-properties}
and Fig.~\ref{TEMimage}) shows the final Py film to be a single,
homogeneous layer rather than two distinct layers. We attribute the
lack of a morphological boundary between the two layers to bombardment
during the second deposition. The sum of the first and second deposition
thicknesses is the nominal Py thickness, $d_{0}$.

\subsection{Gr Properties \label{sub:Gr-Properties}}

We used Raman spectroscopy to measure Gr damage as follows. The Raman
spectrum of pristine Gr has two characteristic peaks, named G and
2D, and defects allow an additional scattering mechanism that adds
a feature named the D peak \citep{Ferrari:2013kx}. The ratio of the
peak intensities for the D and G features, $I_{\textrm{D}}/I_{\textrm{G}}$,
can be used to estimate the number of point defects per unit area
\citep{Cancado:2011dq}.

For our as-grown Gr on Cu, we found a baseline value of $I_{\textrm{D}}/I_{\textrm{G}}=0.11\pm0.03$,
which implies one defect every $\unit[(1300\pm350)]{nm^{2}}$. We
deposited 5~nm to 35~nm of Py at $\theta=90^{\circ}$ on several
samples and found no change in $I_{\textrm{D}}/I_{\textrm{G}}$, demonstrating
that sputtering in this geometry does not damage Gr. The fact that
Gr Raman peaks could be measured through these films indicates the
Py deposited at $\theta=90^{\circ}$ was highly transparent, consistent
with the fact that these samples still showed some Cu color after
deposition. In contrast, Py deposited at $\theta=30^{\circ}$ did
not show any Gr Raman peaks for thickness $\geq\unit[2]{nm}$ and
had a color similar to that of high quality Ni films. A short deposition
at $\theta=30^{\circ}$, lasting $\approx\unit[2]{s}$ and giving
$\approx\unit[0.2]{nm}$, allowed us to measure Gr Raman peaks with
$I_{\textrm{D}}/I_{\textrm{G}}=1.7$, corresponding to one defect
every $\unit[100]{nm^{2}}$. This confirms that Gr is readily damaged
by sputtering in this geometry, with a damage rate of roughly $\unit[0.005]{nm^{-2}\cdot s^{-1}}$.
In fact, this rate is close to the flux of Ar neutrals expected from
the simulations in \citep{Yamamura:1995nx} (see Appendix~C).

We estimated the potential damage to Gr covered by Py using the Stopping
and Range of Ions in Matter (SRIM) Monte Carlo simulation \citep{Ziegler:2010rt}.
Since the substrate in our case is located 23~cm from the target,
only Ar neutrals reflected nearly perpendicular to the target will
reach the substrate, with a maximum energy of $\approx$ 70~eV for
a 430~V discharge \citep{Yamamura:1995nx}. For $10^{7}$ Ar ions
at 70~eV and normal incidence, SRIM predicts no transmission for
5~nm of Py with its nominal density of $\unit[8.7]{g/cm^{3}}$. Adjusting
parameters to reflect the fact that Py deposited at $\theta=90^{\circ}$
is different from nominal, \emph{e.g.,} using a thickness of 2~nm
and nominal density, or a thickness of 5~nm and a density of $\unit[3]{g/cm^{3}}$,
yielded a transmission probability of less than 1 in $10^{7}$. Combining
this with the Ar neutral flux estimated in Appendix~C shows that
damage due to Ar penetration through the $\theta=90^{\circ}$ Py film
is expected to be negligible.

\subsection{Py properties \label{sub:Py-properties}}

In addition to FMR measurements, we used vibrating sample magnetometry
(VSM) to measure Py magnetostatic properties of the same samples presented
in Fig.~\ref{DampingVsThickness}. We also used transmission electron
microscopy (TEM) to characterize Py morphology for separate, similarly
prepared samples.

\begin{figure*}[!]
\begin{centering}
\includegraphics[bb=0bp 0bp 275bp 293bp,width=8.5cm]{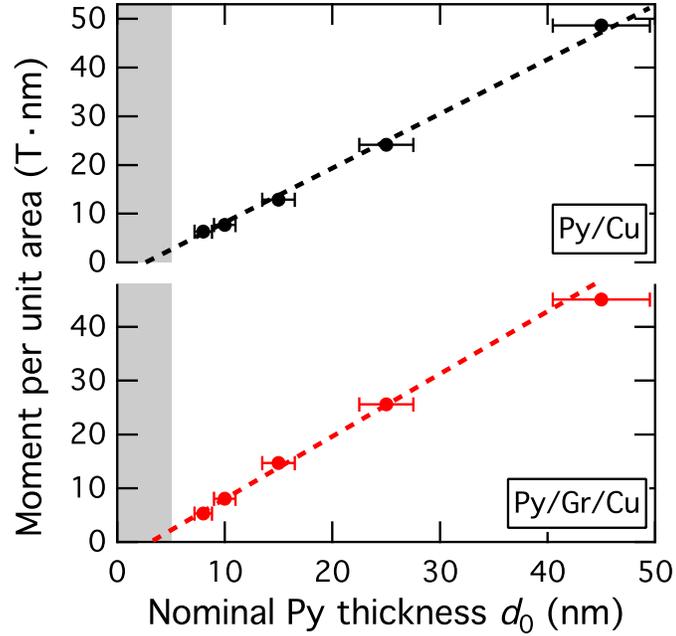}
\par\end{centering}

\centering{}\protect\caption{Magnetic moment per unit area vs nominal Py thickness from VSM measurements.
The gray shaded region indicates the initial 5~nm of Py deposited
at glancing incidence, $\theta=90^{\circ}$. The rest of the Py was
deposited at $\theta=30^{\circ}$. Horizontal error bars reflect an
estimated 10\% uncertainty in deposition rate, and the fits are weighted
accordingly. \label{MomentVsThickness} }
\end{figure*}

For a homogeneous FM film with magnetization $M_{\textrm{s}}$ and
thickness $d_{0}$, the moment per unit area is $\mu_{0}M_{\textrm{s}}d_{\textrm{0}}$.
The total magnetic moment measured by VSM, divided by the area of
each sample, is plotted vs. nominal Py thickness in Fig.~\ref{MomentVsThickness}.
From the slopes of the linear fits in Fig~\ref{MomentVsThickness},
we find $\mu_{0}M_{\mathrm{s}}=\unit[\left(1.1\pm0.1\right)]{T}$
for both Py/Cu and Py/Gr/Cu samples. The nonzero intercepts of the
fits imply the actual thickness of the films is $\approx\unit[3]{nm}$
less than the nominal thickness. This is not surprising, since the
5~nm of Py deposited at $\theta=90^{\circ}$ is porous (see Section~\ref{sub:Gr-Properties})
and some of the Py deposited at $\theta=30^{\circ}$ will fill the
pores instead of adding to the overall film thickness. For the rest
of this paper, we use the actual FM thickness $d_{\mathrm{FM}}$ obtained
by subtracting the intercept values from $d_{0}$.

\begin{figure*}[!]
\begin{centering}
\includegraphics[bb=0bp 0bp 275bp 293bp,width=8.5cm]{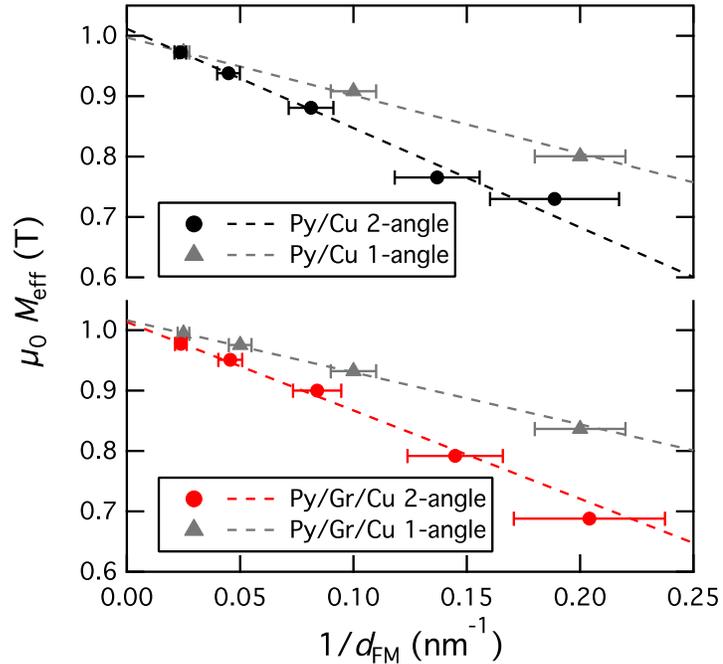}
\par\end{centering}

\centering{}\protect\caption{$M_{\mathrm{eff}}$ vs. $1/d_{\mathrm{FM}}$ with weighted linear
fits for samples deposited at two angles and at a single angle ($\theta=30^{\circ}$).
\label{MeffVsThickness}}
\end{figure*}

Given the actual FM thickness values, we can use $M_{\mathrm{eff}}$
measured by FMR to determine the perpendicular anisotropy of the Py
as follows. Figure~\ref{MeffVsThickness} shows plots of $M_{\mathrm{eff}}$
vs. $1/d_{\mathrm{FM}}$ for the two-angle Py/Cu and Py/Gr/Cu samples,
and for several samples deposited only at $\theta=30^{\circ}$. The
linear dependence for all samples is consistent with a constant interfacial
anisotropy energy $K_{\mathrm{int}}$, for which $M_{\mathrm{eff}}$
is given by

\[
M_{\textrm{eff}}=M_{\textrm{s}}-\frac{2K_{\textrm{int}}}{\mu_{0}M_{\textrm{s}}d_{\textrm{FM}}}.
\]
Thus the intercept of each linear fit in Fig.~\ref{MeffVsThickness}
gives $M_{\textrm{s}}$ and the slope gives $K_{\mathrm{int}}$. We
find $\mu_{0}M_{\mathrm{s}}=\unit[\left(1.01\pm0.02\right)]{T}$ for
both Py/Cu and Py/Gr/Cu samples, consistent with the value obtained
from VSM data. The values of $K_{\mathrm{int}}$, given in Table~\ref{Anisotropy_Energies_Table},
are quite similar for a given deposition method, but are noticeably
larger for two-angle films than for single-angle films. Furthermore,
our values are larger than the value of $K_{\mathrm{int}}\approx1\times10^{-4}\unit{J/m^{2}}$
reported for Py/Cu \citep{Rantschler:2005db} and Py/air \citep{Bailey:1973kq}
interfaces. These differences are likely due to variations is how
Py nucleates and grows on different substrates, for example polycrystalline
Cu compared to our Cu(111), in addition to the different deposition
geometries.

\begin{table}[b]
\begin{centering}
\begin{tabular*}{8.6cm}{@{\extracolsep{\fill}}cc}
\toprule & $K_{\mathrm{int}}$ $\left(\unit{J/m^{2}}\right)$\tabularnewline
\midrule 
Py/Cu, two-angle & $\left(6.6\pm0.6\right)\times10^{-4}$\tabularnewline
Py/Gr/Cu, two-angle & $\left(5.6\pm0.6\right)\times10^{-4}$\tabularnewline
Py/Cu, $\theta=30^{\circ}$ & $\left(3.8\pm0.3\right)\times10^{-4}$\tabularnewline
Py/Gr/Cu, $\theta=30^{\circ}$ & $\left(3.5\pm0.3\right)\times10^{-4}$\tabularnewline\bottomrule
\end{tabular*}
\par\end{centering}

\protect\caption{Anisotropy energies from FMR measurements. \label{Anisotropy_Energies_Table}}
\end{table}

The coercive field measured by VSM is 0.5~mT for Py/Cu and 0.25~mT
for Py/Gr/Cu. Although both values are somewhat larger than the $\approx\unit[0.1]{mT}$
typically found for high-quality Py, similar values have been reported
for Py deposited on Cu(111) single crystals \citep{Rook:1991rz}.
Finally, a 40~nm Py/Cu film deposited at a single angle of $\theta=30^{\circ}$
showed the same value of $\alpha$ ($\approx0.008$) as the two-angle
Py/Cu film with 5~nm at $\theta=90^{\circ}$ and 40~nm at $\theta=30^{\circ}$,
indicating that the intrinsic damping of the Permalloy is not affected
by two-angle deposition. 

Overall, these results show that whatever effects Gr may have on the
film growth process, for our two-angle method it has little or no
effect on the Py magnetic properties.

We prepared samples for cross-sectional TEM imaging using a standard
lift-out technique in a focused ion beam microscope \citep{Giannuzzi:1997kq}.
Following a thinning step using 5~kV Ga\textsuperscript{+} ions,
we performed a final cleaning and thinning using 900~eV Ar\textsuperscript{+}
ions.

\begin{figure}[!h]
\begin{centering}
\includegraphics[width=8.5cm]{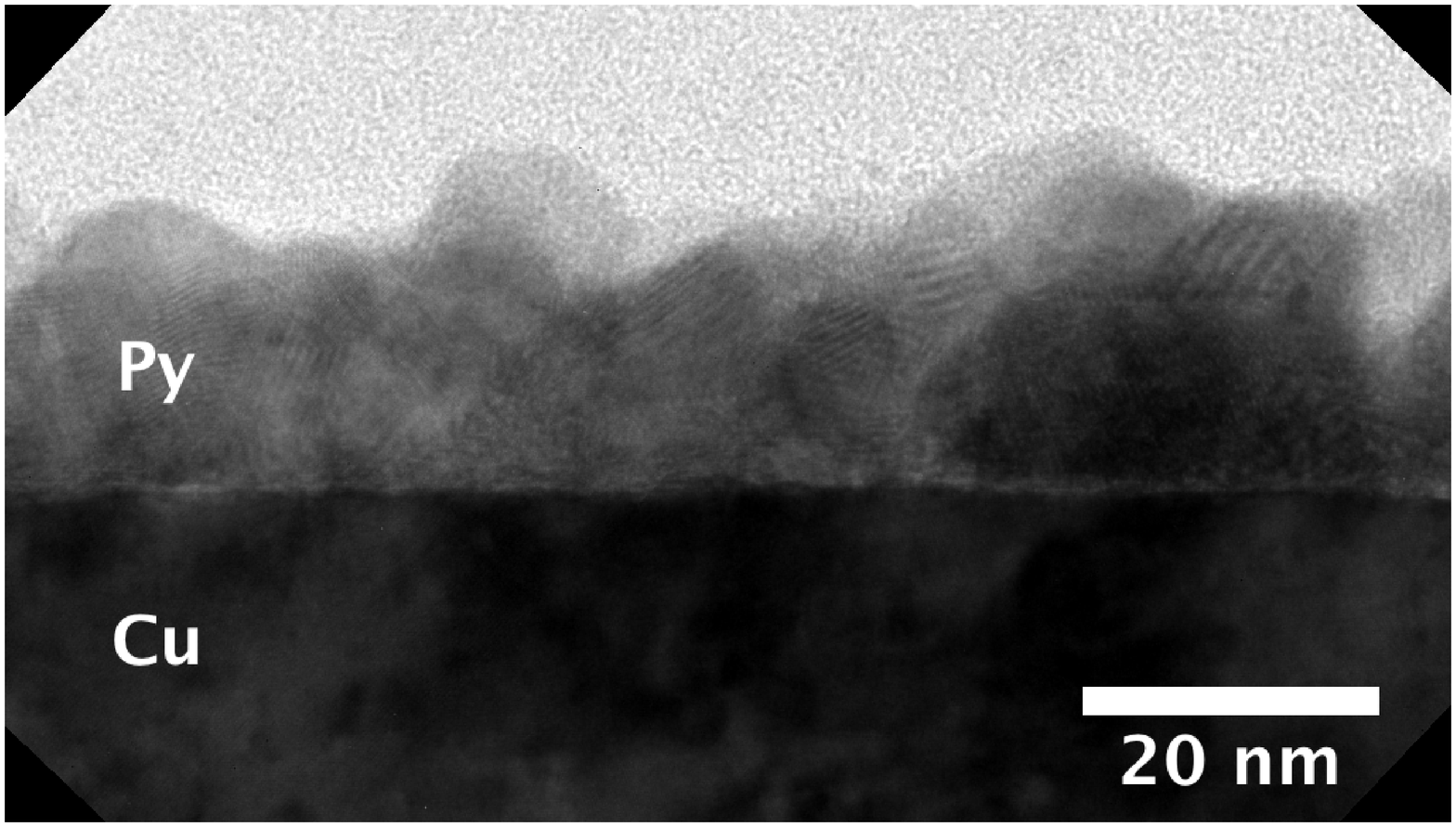}
\par\end{centering}

\centering{}\protect\caption{TEM cross-section image of a Py/Gr/Cu sample with 10~nm deposited
at $\theta=90^{\circ}$ and 10~nm deposited at $\theta=30^{\circ}$.
\label{TEMimage}}
\end{figure}

An image of a two-angle Py film deposited on Gr/Cu, 10~nm at $\theta=90^{\circ}$
and 10~nm at $\theta=30^{\circ}$, is shown in Fig.~\ref{TEMimage}.
While the film is quite rough, it is continuous and, importantly,
shows neither internal voids nor an apparent boundary between the
two Py layers. This suggests that any holes in the first layer are
filled in during the second deposition. The Py grains are fairly round
and selected area diffraction patterns (not shown) indicate a modest
(111) texture.

\section{Discussion}

Since our co-deposited Py/Cu and Py/Gr/Cu films have the same magnetization,
similar perpendicular anisotropies, and nearly the same damping for
films 15~nm and thicker, the dramatic difference in damping for thinner
films (see Fig.~\ref{DampingVsThickness}) must be attributed to
the Gr interlayer. We do not expect other effects such as surface
oxidation and surface roughness to be substantially different between
the two types of samples. This leads to the surprising conclusion
that a single layer of C atoms can block the transfer of spins between
two metals. Before offering an explanation for this effect, we first
compare our results with other FMR measurements of Py films deposited
on Gr.

In \citep{Patra:2012fk} and \citep{Singh:2015fe}, Gr grown by CVD
on Cu foils was transferred to insulating or semiconducting substrates
and Py was deposited by thermal evaporation to give Py/substrate and
Py/Gr/substrate samples. Only a single Py thickness, 14~nm, was used
for all samples in these two studies. In \citep{Patra:2012fk}, for
samples where the Py covered the entire Gr area, the Py damping was
1.9~times larger when Gr was present. This was interpreted as evidence
for strong spin scattering in Gr itself, which would be contrary to
all expectations. However, $M_{\textrm{eff}}$ was also different
for these samples (30\% lower with Gr present) which suggests the
difference in damping could be due to differences in the Py itself
rather than the presence of Gr. In \citep{Singh:2015fe}, the same
group made Py/Gr samples where the Gr extended $\approx\unit[10]{\mu m}$
past two edges of the Py film. Damping for these sample was 1.1~times
larger than for similar samples where the Py covered the entire Gr
area. This small change in damping cannot be reliably attributed to
spin pumping without further measurements to rule out changes in the
Py. Unfortunately, no independent measurements of $M_{\textrm{s}}$
or $H_{\textrm{k}}^{\bot}$ were reported in \citep{Patra:2012fk}
or \citep{Singh:2015fe}. Further evidence of the need for caution
comes from a study of 14~nm thick Co films deposited by evaporation
onto Gr/SiO$_{2}$ and SiO$_{2}$ substrates \citep{Berger:2014xe}.
Careful characterization by FMR, magnetometry, magnetic force microscopy,
and Kerr microscopy revealed such clear magnetic differences between
the samples that the authors, commendably, declined to draw any conclusions
about spin pumping into Gr.

In \citep{Tang:2013zr}, Py and Pd were evaporated onto a Gr/SiO$_{2}$
substrate and patterned into stripes separated by about $\unit[1]{\mu m}$
of unpatterned Gr. When the Py was driven to precess by an $\unit[9.62]{GHz}$
magnetic field, a voltage was detected in the Pd stripe that behaved
as expected if a spin current was pumped from the Py into the Gr channel
and then generated an inverse spin Hall voltage after diffusing to
the Pd. FMR measurements at this same frequency showed the linewidth
for Py/Gr/SiO$_{2}$ was 0.5~mT larger than for Py/SiO$_{2}$, for
25~nm thick Py. This difference was attributed to spin pumping from
Py to Gr, but since $\Delta H_{0}$ was not measured the difference
could equally be due to differences in the Py itself. The fact that
$\Delta H_{0}$ for our 25~nm Py films differed by 0.9~mT, while
$\alpha$ differed by only 1\%, shows that FMR linewidth at a single
frequency does not yield a reliable estimate for damping or spin pumping.

The limitations of these previous experiments can be avoided by varying
both FMR frequency and FM thickness over a wide range. The frequency
range reveals what fraction of FMR linewidth is due to damping ($\alpha$)
as opposed to inhomogeneity ($\Delta H_{0}$). The plot of $\alpha$
vs. $d_{\textrm{FM}}$ allows a clear separation of the intrinsic
and spin pumping contributions to $\alpha$ at each thickness.

\begin{figure}[!h]
\begin{centering}
\includegraphics[width=8.5cm]{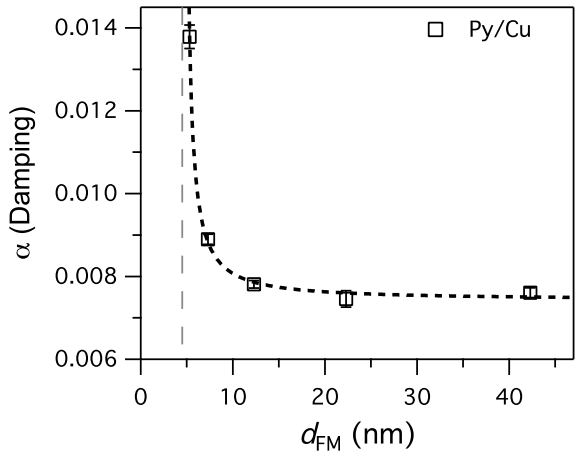}
\par\end{centering}

\centering{}\protect\caption{Fit to damping vs. thickness for Py/Cu samples using Eq.~\ref{eq:damping_Geff}.
\label{SpinPumpingFit}}
\end{figure}

For our Py/Cu samples, the spin pumping expression in Eq.~\ref{eq:damping_Geff}
can be fit to the data shown in Fig.~\ref{DampingVsThickness} and
reproduced in Fig.~\ref{SpinPumpingFit}. For this fit we used the
value of $M_{\textrm{s}}^{\textrm{}}$ calculated from the data in
Fig. \ref{MeffVsThickness}. We also added an adjustable thickness
$t_{\textrm{d}}^{\textrm{SP}}$, representing a Py ``dead layer''
that does not contribute to spin pumping. Finally, we added an adjustable
intrinsic damping $\alpha_{0}$ to account for non-zero damping at
large thickness. The fit then yields $\alpha_{0}=0.0074\pm0.0001$,
$G_{\mathrm{eff}}=\unit[(8.6\pm1.6)\times10^{13}]{\Omega^{-1}\cdot m^{-2}}$,
and $t_{\textrm{d}}^{\textrm{SP}}=\unit[4.7\pm0.7]{nm}$, and is shown
by the dashed curve in Fig.~\ref{SpinPumpingFit}. The uncertainty
in $t_{\textrm{d}}^{\textrm{SP}}$ is largely due to uncertainty in
the thickness of the thinnest film.

As described in Appendix A, we use a model for $G_{\mathrm{eff}}$
that combines an interfacial spin mixing conductance $G_{\uparrow\downarrow}$
and an external conductance $G_{\mathrm{ext}}$ as resistors in series,
i.e., 
\begin{equation}
\frac{1}{G_{\textrm{eff}}}\equiv\frac{1}{G_{\uparrow\downarrow}}+\frac{1}{G_{\textrm{ext}}}.\label{eq:Geff}
\end{equation}
The interfacial contribution is of order $\unit[10^{15}]{\Omega^{-1}\cdot m^{-2}}$
(see Appendix A), so $G_{\mathrm{eff}}$ in our case is dominated
by $G_{\mathrm{ext}}$ and insensitive to the value of $G_{\uparrow\downarrow}$.
We can therefore neglect $G_{\uparrow\downarrow}$ and, using the
expression for $G_{\mathrm{ext}}$ given in Appendix A and a bulk
conductivity for Cu of $\sigma=\unit[5.8\times10^{7}]{\Omega^{-1}\cdot m^{-1}}$,
we infer a spin diffusion length in Cu of $\lambda_{\mathrm{s}}=\unit[310\pm60]{nm}$.
This value is roughly consistent with the various room temperature
values compiled in \citep{Bass:2007uq}.

While we are unable to find prior measurements of spin pumping dead
layers at Py/Cu interfaces, measurements on other FM/NM systems show
much thinner dead layers, $\sim\unit[1]{nm}$ \citep{Boone:2013ly}.
This could be due to differences in the Py/Gr interface due to our
unusual deposition conditions. Additionally, our 1-D spin pumping
model does not account for lateral sample inhomogeneity, such as that
from roughness observed in Fig.~\ref{TEMimage}, but the effect of
in-plane spin currents on our parameters is unknown. 

A likely explanation for the dramatic suppression of spin pumping
by a single Gr layer is the electrical contact resistance between
Gr and Cu. Measurements for several metals deposited on Gr gave a
value of $1/G_{\textrm{c}}\approx\unit[10^{-9}]{\Omega\cdot m^{2}}$
for Cu and values of $\approx\unit[2\times10^{-11}]{\Omega\cdot m^{2}}$
for Ni, Ti, Pd, and Pt \citep{Robinson:2011uq}. Adding this term
to Eq.~\ref{eq:Geff}, we expect $G_{\textrm{eff}}\approx G_{\textrm{c}}\approx\unit[10^{9}]{\Omega^{-1}\cdot m^{-2}}$
for Py/Gr/Cu, which is nearly 5 orders of magnitude smaller than $G_{\mathrm{eff}}$
for Py/Cu. Thus the Gr/Cu interface presents a large barrier to spin
flow, reducing the net flow of spin current away from the Py to the
point where even our thinnest films show no enhanced damping.

Even the lowest metal-Gr contact resistances are much larger than
$1/G_{\uparrow\downarrow}$, which raises the question of whether
spin pumping into Gr is possible under any conditions. Although the
values of $G_{\uparrow\downarrow}$ and $\lambda_{\mathrm{s}}$ inferred
in \citep{Tang:2013zr} depend entirely on a questionable value for
$\Delta\alpha_{\textrm{sp}}$, the voltage detected at the Pd in those
experiments is strong evidence that spins did flow from the Py to
the Pd. Given this evidence, we believe it is likely that in our Py/Gr/Cu
samples there is significant spin pumping into the Gr, but also a
large spin accumulation because of the barrier at the Gr/Cu interface
and therefore a large diffusion of spins back into the Py. A quantitative
understanding of this situation, in which the same monolayer of C
atoms participates in both the Py/Gr and the Gr/Cu interfaces, is
a rich topic for further investigation.

\section{Conclusion}

In summary, we prepared Py/Cu and Py/Gr/Cu samples with varying Py
thickness $d_{\textrm{FM}}$, using a two-angle sputtering method
and carefully verifying that neither Gr nor Py were degraded. We used
FMR to measure damping $\alpha$ for both series of samples. The Py/Cu
samples showed an increase in $\alpha$ with decreasing $d_{\textrm{FM}}$,
as expected from spin pumping. Fitting the standard model of spin
pumping showed $G_{\mathrm{eff}}$ was dominated by $G_{\mathrm{ext}}$
of the Cu rather than the interfacial conductance $G_{\uparrow\downarrow}$,
which precludes determining a value for the latter. The Py/Gr/Cu samples,
despite having the same Py properties, showed no change in damping
with thickness. This implies a strong suppression of spin flow away
from the Py, which we attribute to the large contact resistance at
the Gr/Cu interface.

Our results are consistent with the evidence for spin pumping from
Py into Gr reported in \citep{Tang:2013zr}, where an inverse spin
Hall voltage was detected at a separate Pd electrode. However, we
emphasize that quantitative analysis of spin pumping data requires
independent characterization of magnetic properties and a consideration
of the effect of all relevant conductances, not only $G_{\uparrow\downarrow}$.
Our results are not consistent with the interpretation offered in
\citep{Patra:2012fk} that Gr itself is a strong spin absorber. We
expect that magnetometry and/or a thickness series would show that
differences in Py properties, rather than spin pumping, are responsible
for the change in damping in this case.

\section{{\normalsize{}Methods}}

\subsection{Cu film deposition and Gr CVD}

Epitaxial, crystalline Cu(111) thin films on sapphire were made as
described in \citep{Miller:2013fk}. In brief, 500~nm of Cu was sputtered
onto 50~mm wafers of $\alpha$-Al$_{2}$O$_{3}$(0001) held at $\unit[65]{^{\circ}C}$.
During the annealing that preceded Gr growth (see below), secondary
grain growth resulted in exclusively Cu(111) grains $\gtrsim\unit[2]{cm}$
across. These large grains prevent dewetting of the Cu film during
Gr CVD at higher temperatures \citep{Miller:2013fk}. One Cu/Al$_{2}$O$_{3}$
wafer was removed after annealing only and used for Py deposition
directly on Cu, while a second wafer continued through the Gr growth
step.

Gr CVD was performed in a hot-wall quartz tube furnace with a diameter
of 76~mm. After evacuating the tube with a dry pump to a pressure
of $\unit[0.67]{Pa}$ (5~mTorr), the rest of the process was performed
at a total pressure of $\unit[80]{kPa}$ (600~Torr). The wafer was
heated in a flow of 7~sccm H$_{2}$ and 2100~sccm Ar (H$_{2}$ partial
pressure of $\unit[0.3]{kPa}$ (2~Torr)), initially ramping to $\unit[920]{^{\circ}C}$
in about 15~min and then moving to $\unit[1000]{^{\circ}C}$ at $\unit[2]{^{\circ}C/min}$
to allow complete Cu grain growth. The temperature was then increased
to $1060{}^{\circ}$C and the gas flows were changed to 58~sccm H$_{2}$
and 3600~sccm Ar (H$_{2}$ partial pressure of $\unit[1.3]{kPa}$
(9.5~Torr)). After 10~min under these conditions, Gr growth was
initiated by adding 18~sccm of 0.2\% CH$_{4}$ in Ar (CH$_{4}$ partial
pressure of $\unit[0.8]{Pa}$ (6~mTorr)). Growth conditions were
maintained for 3~h before rapid cooling by opening the furnace lid.
The flow of CH$_{4}$ was stopped below $400{}^{\circ}$C and the
flow of H$_{2}$ was stopped below $\unit[100]{^{\circ}C}$.

The resulting Gr films were characterized by optical microscopy to
check for completeness of growth and Raman spectroscopy to confirm
their quality. For shorter growth times, for which Gr coverage was
incomplete, we found this growth recipe consistently gave compact
hexagonal Gr growth domains with nucleation sites separated by $\approx\unit[100]{\text{\textmu m}}$.

Both wafers were cut into 6 mm $\times$ 8 mm chips using a diamond
scribe.

\subsection{Py sputter deposition}

For each thickness, Py was deposited simultaneously onto bare Cu and
Gr/Cu samples. To remove Cu oxide and adventitious contamination,
samples were soaked in glacial acetic acid for 1~min and rinsed in
deionized water just prior to being placed in the load lock of a deposition
chamber with a base pressure $\approx\unit[\unit[4\times10^{-6}]{Pa}]{}$
($\unit[3\times10^{-8}]{Torr}$). The samples were heated to $\approx\unit[200]{^{\circ}C}$
to remove residual water and then cooled in vacuum to $\approx\unit[50]{^{\circ}C}$
before deposition. Depositions at both $\theta=90^{\circ}$ and $\theta=30^{\circ}$
were performed with the source running at 200~W and $\approx\unit[430]{V}$,
in 1.1~mTorr of Ar gas, and with a target-to-sample distance of 23~cm.
The deposition rates, measured using atomic force microscopy for $\approx\unit[10]{nm}$
thick films, were $\unit[0.03]{nm/s}$ for $\theta=90^{\circ}$ and
$\unit[0.09]{nm/s}$ for $\theta=30^{\circ}$.

\subsection{Magnetometry}

The VSM measurements were performed at room temperature with the sample
mounted in a plastic straw, using a frequency of 40~Hz, and an oscillation
amplitude of 2~mm. The magnetic field, applied in the plane of the
sample, was swept over a range of $\mu_{0}H=\pm$4~T to allow a linear
fit to the diamagnetic background, due primarily to the Al$_{2}$O$_{3}$
substrate. The moment of the Py layer was taken as the $H=0$ intercept
value of a linear fit from 1~T to 4~T.

\subsection{Raman spectroscopy\label{sub:Raman-sect}}

We performed Raman spectroscopy in a home-built system using a 532~nm
diode laser with a power of $\approx\unit[3.5]{mW}$ and a spot size
$<\unit[10]{\text{\ensuremath{\mu}m}}$. Peak fitting was done after
subtracting a smooth fit to the surrounding background from Cu fluorescence.

\section*{{\normalsize{}Acknowledgments}}

The authors thank Justin Shaw for valuable comments on interfacial
anisotropy in Py, Bill Rippard and Justin Shaw for help in maintaining
the sputter deposition chamber, and Ben Derby for help in sample preparation.
Optical microscopy and TEM were done in NIST's Precision Imaging Facility.

\section*{{\normalsize{}Author contributions}}

MWK, TJS, and WJG designed the experiments. WJG made the Gr/Cu substrates,
deposited the Py films, and performed VSM and Raman measurements.
HTN performed FMR measurements and TJS peformed FMR data analysis.
ACD performed TEM sample preparation and imaging. MWK developed the
two-angle sputtering method. MWK and WJG wrote the manuscript. All
authors contributed to interpretation of the data and commented on
the manuscript.

\section*{APPENDIX A: Model for Spin Pumping in Multi-Layer Structures}

\begin{figure}[!th]
\centering{}\includegraphics[width=8.5cm, clip=true]{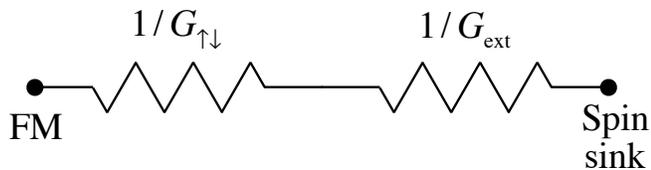}\protect\caption{Series resistor model for spin transport between a FM and a NM spin
sink. \label{Series_Resistor_Model}}
\end{figure}

Following \citep{Boone:2013ly}, we model the spin transport of a
FM layer in contact with one or more NM layers as an electrical circuit
of resistors in series. Fig.~\ref{Series_Resistor_Model} shows the
simplest case of one interfacial resistance, $1/G_{\uparrow\downarrow}$,
and one external resistance, $1/G_{\mathrm{ext}}$, The effective
conductance for spin current flowing from the FM to the spin sink
is simply

\[
\frac{1}{G_{\textrm{eff}}}\equiv\frac{1}{G_{\uparrow\downarrow}}+\frac{1}{G_{\textrm{ext}}}.
\]
Since these conductances describe spin transport perpendicular to
the FM/NM interface, they are defined per unit area of the interface
and have units of $\unit{\textrm{\ensuremath{\Omega}}^{-1}\cdot m^{-2}}$.
In the full theory of spin pumping \citep{Brataas:2014ys}, $G_{\uparrow\downarrow}$
is a complex quantity whose real part affects damping and whose imaginary
part affects the FM precession frequency. For conventional FM metals
such as Ni, Fe, and Co, the effect on frequency is negligible \citep{Zwierzycki:2005qv}
and $G_{\uparrow\downarrow}$ is commonly used to refer only to the
real part.

The interfacial conductance $G_{\uparrow\downarrow}$, often called
the ``spin-mixing'' conductance, determines how readily spins can
move across the FM/NM interface. This process is governed by the exchange
interaction, which operates over a length scale of a few nanometers.
For metallic interfaces, the theoretical upper bound on $G_{\uparrow\downarrow}$
\citep{Bauer:2003dz} is the Sharvin conductance \citep{Sharvin:1965rz}
(for a single spin direction and per unit area), corresponding to
perfect transmission for every conductance channel at the interface,
\[
G_{\textrm{S}}=\frac{e^{2}}{h}\frac{k_{\textrm{F}}^{2}}{4\pi},
\]
where $k_{\textrm{F}}$ is the Fermi wavevector. Taking Cu as a typical
metal, with $k_{\textrm{F}}=\unit[1.36\times10^{10}]{m^{-1}}$, we
find $G_{\textrm{S}}\approx\unit[0.5\times10^{15}]{\Omega^{-1}\cdot m^{-2}}$.
Both experimental and theoretical results for Py/Cu interfaces (and
several other FM/NM combinations) are consistent with this limit \citep{Tserkovnyak:2002tg,Xia:2002dk,Bauer:2003dz,Gerrits:2006qy,Shaw:2012wu},
with $G_{\uparrow\downarrow}$ in the range of $\unit[(0.5\pm0.1)\times10^{15}]{\Omega^{-1}\cdot m^{-2}}$.

The external conductance $G_{\mathrm{ext}}$ determines how readily
spins flow away from the interface once they have entered the NM material.
For a single NM layer of thickness $d_{\textrm{NM}}$, bulk electrical
conductivity $\sigma$, and spin diffusion length $\lambda_{\textrm{s}}$,
$G_{\mathrm{ext}}$ is given in \citep{Boone:2013ly} as

\[
G_{\mathrm{ext}}=\frac{1}{2}\frac{\sigma}{\lambda_{\textrm{s}}}\tanh\frac{d_{\textrm{NM}}}{\lambda_{\textrm{s}}}.
\]
For $d_{\textrm{NM}}\gtrsim2\lambda_{\textrm{s}}$, which is often
the case, the external spin conductance is simply determined by the
ratio of conductivity and spin diffusion length, $G_{\mathrm{ext}}\approx\sigma/2\lambda_{\textrm{s}}$.

In the limit of $G_{\mathrm{ext}}\gg G_{\uparrow\downarrow}$, when
spins diffuse rapidly away from the interface, $G_{\mathrm{eff}}\approx G_{\uparrow\downarrow}$
and a measurement of $\Delta\alpha_{\textrm{sp}}$ can be used to
directly infer a value of $G_{\uparrow\downarrow}$ from Eq.~\ref{eq:damping_Geff}.
This limit applies to NM materials with large $\sigma$ and/or small
$\lambda_{\textrm{s}}$, such as Pt and Pd \citep{Boone:2013ly}.
However, other materials such as Ta and Cu \citep{Boone:2013ly} are
in the opposite limit, where $G_{\mathrm{eff}}\approx G_{\textrm{ext}}$
and $\Delta\alpha_{\textrm{sp}}$ is not sensitive to $G_{\uparrow\downarrow}$.

\section*{APPENDIX B: Full FMR Results}

\begin{figure}[htb!]
\begin{centering}
\includegraphics[width=8.5cm]{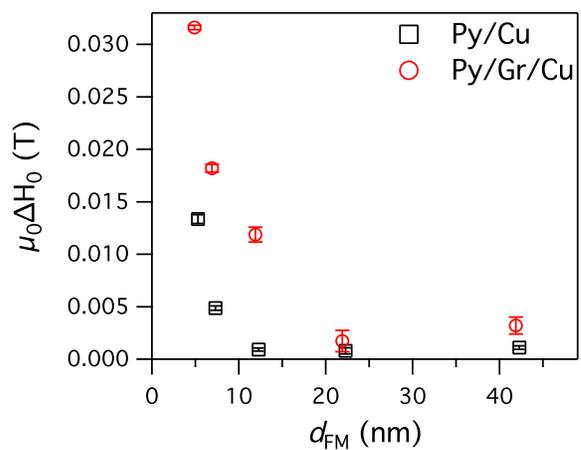}
\par\end{centering}

\centering{}\protect\caption{Plot of $\Delta H_0$ vs Py thickness.\label{FMRdata-DeltaH_0}}
\end{figure}

\begin{figure}[htb!]
\begin{centering}
\includegraphics[width=8.5cm]{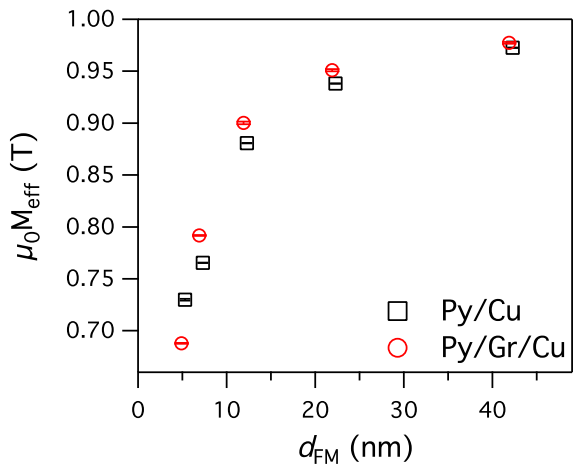}
\par\end{centering}

\centering{}\protect\caption{Plot of $M_\textrm{eff}$ vs Py thickness.\label{FMRdata-M_eff}}
\end{figure}

\begin{figure}[htb!]
\begin{centering}
\includegraphics[width=8.5cm]{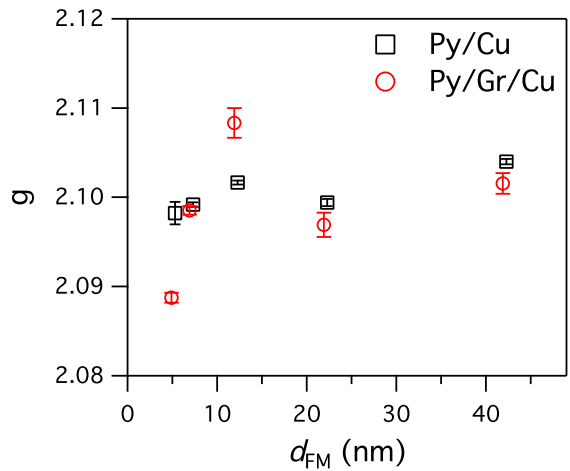}
\par\end{centering}

\centering{}\protect\caption{Plot of $g$ vs Py thickness.\label{FMRdata-g}}
\end{figure}

Table \ref{FMRtable1} shows all parameters obtained by fitting the
FMR data for each sample. These results are plotted in Figs. \ref{FMRdata-DeltaH_0},
\ref{FMRdata-M_eff}, and \ref{FMRdata-g}. Values of $g$ come from
fits to Equation \ref{eq:H_resonance}, where $\gamma=\frac{g\mu_{\mathrm{B}}}{\hbar}$. 

\begin{table*}[tbh]
\begin{centering}
\begin{tabular*}{1\textwidth}{@{\extracolsep{\fill}}cccccc}
\toprule & \multicolumn{1}{c}{$d_{\textrm{0}}$ (nm)} & $\mu_{0}M_{\textrm{eff}}$ (T)  & $g$ & $\mu_{0}\Delta H_{0}$ (mT) & $\alpha$\tabularnewline
\midrule
Cu+Gr & 5+3 & 0.6879 & 2.089 & 31.6 & 0.0073$\pm$0.0001\tabularnewline
 & 5+3 & 0.7918 & 2.098 & 18.2 & 0.0077$\pm$0.0002\tabularnewline
 & 5+10 & 0.9002 & 2.108 & 11.9 & 0.0072$\pm$0.0005\tabularnewline
 & 5+20 & 0.9509 & 2.097 & 1.7 & 0.0084$\pm$0.0007\tabularnewline
 & 5+40 & 0.9772 & 2.102 & 3.2 & 0.0089$\pm$0.0005\tabularnewline
\midrule
Cu & 5+3 & 0.7298 & 2.098 & 13.4 & 0.0138$\pm$0.0003\tabularnewline
 & 5+5 & 0.7654 & 2.099 & 4.9 & 0.0089$\pm$0.0001\tabularnewline
 & 5+10 & 0.8807 & 2.102 & 0.9 & 0.0078$\pm$0.0001\tabularnewline
 & 5+20 & 0.9381 & 2.099 & 0.8 & 0.0075$\pm$0.0002\tabularnewline
 & 5+40 & 0.9725 & 2.104 & 1.1 & 0.0076$\pm$0.0001 \tabularnewline\bottomrule
 &  &  &  &  & \tabularnewline
\end{tabular*}
\par\end{centering}

\centering{}\protect\caption{Parameters from FMR fits of $H_{\textrm{res}}(f)$ and $\Delta H(f)$.
\label{FMRtable1} }
\end{table*}

\section*{APPENDIX C: Graphene Damage Estimates}

\subsection*{Expected Damage Rate from Ar Neutrals}

The simulation results in \citep{Yamamura:1995nx} can be applied
to our sputtering conditions to estimate the flux of Ar neutrals arriving
at the substrate with sufficient energy to damage the Gr. In particular,
the results in Fig.~7 of \citep{Yamamura:1995nx} for 400 eV Ar\textsuperscript{+}
ions hitting a Cu target are a good estimate for our case of 430 eV
Ar\textsuperscript{+} hitting a Ni$_{80}$Fe$_{20}$ target. The
probability of an Ar neutral reflected normal to the target and with
sufficient energy to damage Gr is approximately $2\times10^{-4}$
per unit solid angle for each Ar\textsuperscript{+} ion that hits
the target. Our sputtering source operates at a current of 0.47~A,
corresponding to $2.9\times10^{18}$ Ar\textsuperscript{+} ions per
second. Our $\unit[6]{mm}\times\unit[8]{mm}$ substrate, oriented
at $\theta=30^{\circ}$ and at a distance of 230~mm from the target,
subtends a solid angle of $\unit[7.9\times10^{-4}]{sr}$. The result
in Fig.~21 of \citep{Yamamura:1995nx} shows that there is negligible
scattering of Ar neutrals for our pressure-distance product of $\unit[35]{Pa\cdot mm}$.
This yields a flux of damaging Ar neutrals at our substrate of about
$\unit[0.009]{nm^{-2}\cdot s^{-1}}$, remarkably close to the value
of $\unit[0.005]{nm^{-2}\cdot s^{-1}}$ inferred from the Raman spectrum
of the 2~s exposure described in the main text.

\subsection*{Raman Measurement of Gr Damage vs. Angle}

\begin{figure}[htb]
\begin{centering}
\includegraphics[width=8.5cm]{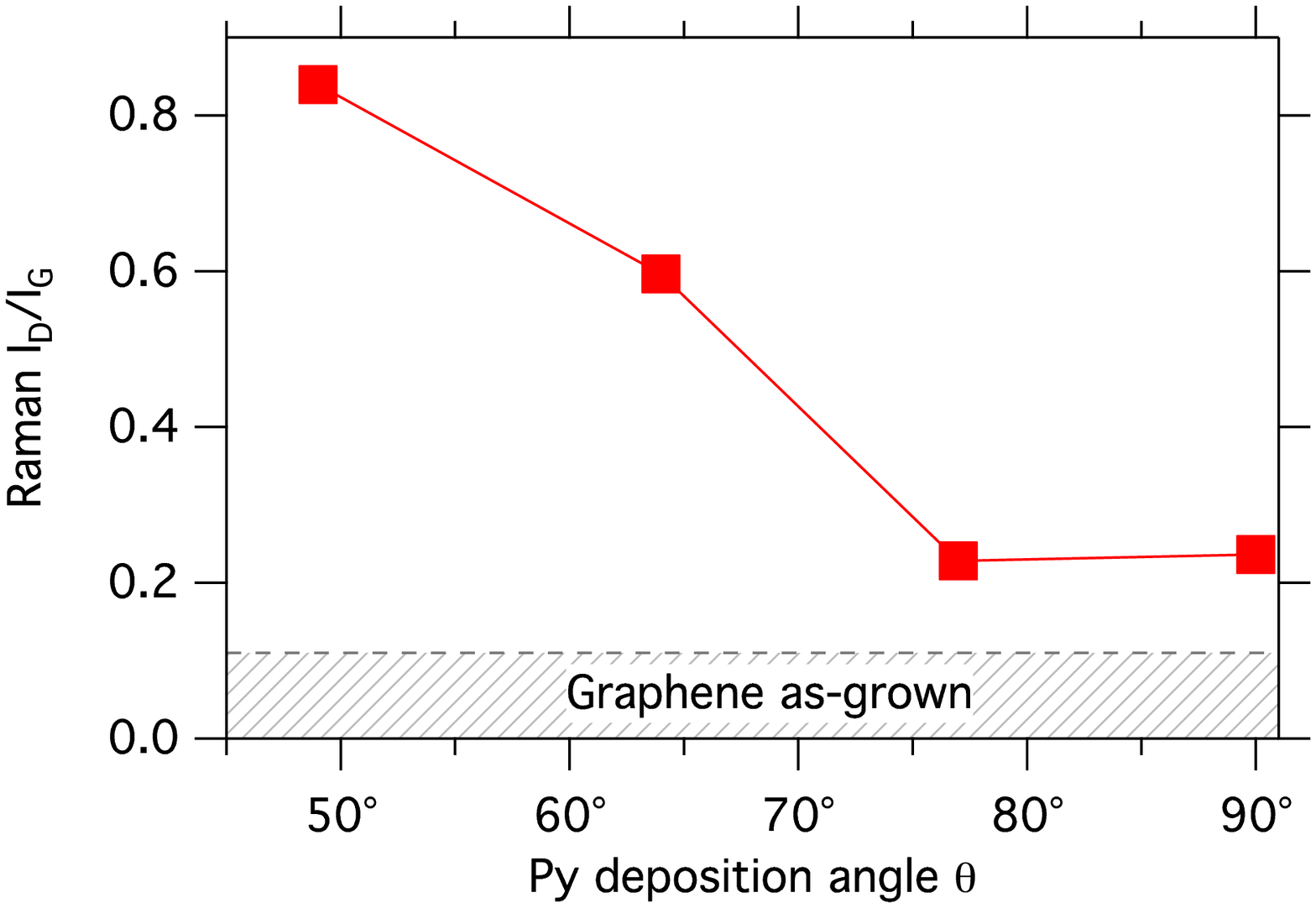}
\par\end{centering}

\begin{centering}
\includegraphics[width=8.5cm]{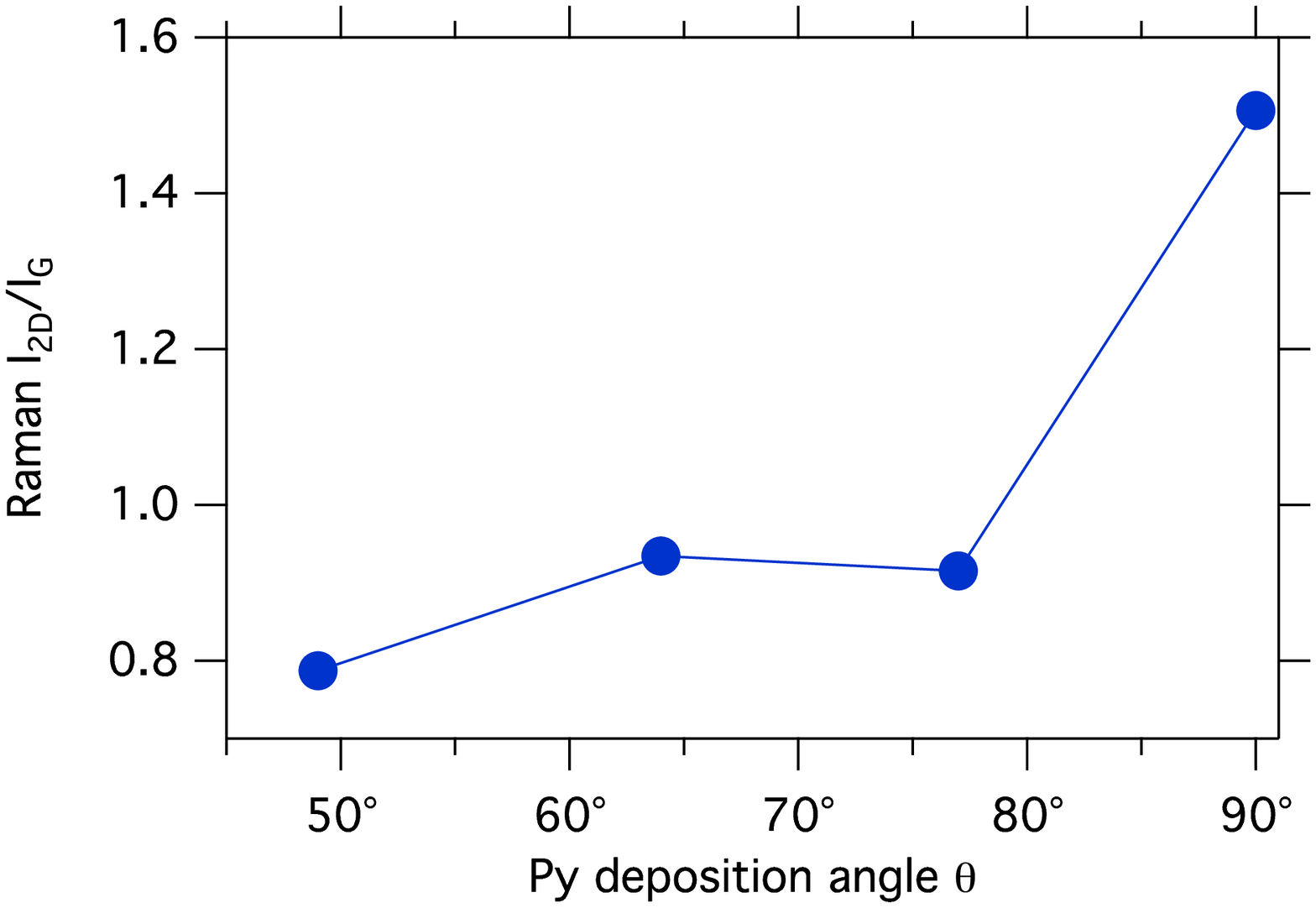}
\par\end{centering}

\centering{}\protect\caption{Raman peak ratios $I_{\textrm{D}}/I_{\textrm{G}}$ and $I_{2\textrm{D}}/I_{\textrm{G}}$
vs Py angle of incidence $\theta$. \label{RamanAngle}}
\end{figure}

In order to evaluate Gr damage as a function of sputtering angle,
we deposited $\approx\unit[10]{nm}$ of Py at several angles between
$\theta=30^{\circ}$ and $\theta=90^{\circ}$ and made Raman measurements.
The results for $I_{\textrm{D}}/I_{\textrm{G}}$ and $I_{2\textrm{D}}/I_{\textrm{G}}$
are shown in Fig.~\ref{RamanAngle}.

\newpage

\bibliographystyle{aipnum4-1}

\end{document}